\documentclass[12pt,a4paper,english]{article}

\usepackage[T1]{fontenc}
\usepackage[latin9]{inputenc}
\usepackage{amsmath}
\usepackage{setspace}
\usepackage{amssymb}

\makeatletter

\providecommand{\tabularnewline}{\\}

\newcommand{\lyxaddress}[1]{
\par {\raggedright #1
\vspace{1.4em}
\noindent\par}
}

\usepackage{babel}
\makeatother

\begin{document}

\title{Generalized Octonion Electrodynamics}

\author{B. C. Chanyal, P. S. Bisht and O. P. S. Negi}

\maketitle
\begin{singlespace}

\lyxaddress{\begin{center}
Department of Physics\\
Kumaun University\\
S. S. J. Campus\\
 Almora -263601 (U.K.) India
\par\end{center}}
\end{singlespace}

\lyxaddress{\begin{center}
Email:- bcchanyal@gmail.com\\
~~~~~~~~~~~~~ps\_bisht123@rediffmail.com\\
~~~~~~~ops\_negi@yahoo.co.in
\par\end{center}}

\begin{abstract}
We have made an attempt to reformulate the generalized field equation
of dyons in terms of octonion variables. Octonion forms of generalized
potential and current equations are discussed in consistent manner.
It has been shown that due to the non associativity of octonion variables
it is necessary to impose certain constraints to describe generalized
octonion electrodynamics in manifestly covariant and consistent manner.
\end{abstract}

\section{Introduction}

~~~~There has been a revival in the formulation of natural laws
so that there exists \cite{key-1} four-division algebras consisting
the algebra of real numbers ($\mathbb{R}$), complex numbers ($\mathbb{C}$),
quaternions ($\mathbb{H}$) and Octonions ($\mathcal{O}$). All four
algebra's are alternative with totally anti symmetric associators.
Quaternions \cite{key-2,key-3} were very first example of hyper complex
numbers have been widely used \cite{key-4,key-5,key-6,key-7,key-8,key-9,key-10}
to the various applications of mathematics and physics. Since octonions
\cite{key-11} share with complex numbers and quaternions, many attractive
mathematical properties, one might except that they would be equally
as useful as others. Octonion \cite{key-11} analysis has been widely
discussed by Baez \cite{key-12}. It has also played an important
role in the context of various physical problems \cite{key-13,key-14,key-15,key-16}
of higher dimensional supersymmetry, super gravity and super strings
etc. In recent years, it has also drawn interests of many \cite{key-17,key-18,key-19,key-20}
towards the developments of wave equation and octonion form of Maxwell's
equations. We \cite{key-21,key-22} have also studied octonion electrodynamics,
dyonic field equation and octonion gauge analyticity of dyons consistently
and obtained the corresponding field equations (Maxwell's equations)
and equation of motion in compact and simpler formulation. Extending
our previous results \cite{key-22}, in this paper, we have made an
attempt to reformulate the generalized field equation of dyons in
terms of octonion variables. Octonion forms of generalized potential
and current equations are discussed in consistent manner. It has been
shown that due to the non associativity of octonion variables it is
necessary to impose certain constraints to describe generalized octonion
electrodynamics in manifestly covariant and consistent manner. We
have obtained the generalized Dirac-Maxwell's (GDM) equations of dyons
from octonion wave equation in simple, compact and consistent manner.
It has been shown that the octonion variable of dyons reproduces the
dynamics of electric (magnetic) charge in the absence of magnetic
(electric) charge.

\section{Octonion Definition}

~~~~~~~An octonion $x$ is expressed \cite{key-21,key-22}
as a set of eight real numbers

\begin{eqnarray}
x=(x_{0},\, x_{1},....,\, x_{7}) & = & x_{0}e_{0}+x_{1}e_{1}+x_{2}e_{2}+x_{3}e_{3}+x_{4}e_{4}+x_{5}e_{5}+x_{6}e_{6}+x_{7}e_{7}\nonumber \\
= & x_{0}e_{0} & +\sum_{A=1}^{7}x_{A}e_{A}\,\,\,\,\,\,\,\,\,\,\,(A=1,2,.....,7)\label{eq:1}\end{eqnarray}
where $e_{A}(A=1,2,.....,7)$ are imaginary octonion units and $e_{0}$
is the multiplicative unit element. The octet $(e_{0},e_{1},e_{2},e_{3},e_{4},e_{5},e_{6},e_{7})$
is known as the octonion basis and its elements satisfy the following
multiplication rules

\begin{eqnarray}
e_{0}=1,\,\,\,\, & e_{0}e_{A}=e_{A}e_{0}=e_{A}\,\,\, & e_{A}e_{B}=-\delta_{AB}e_{0}+f_{ABC}\, e_{C}.\,\,\,\,\,(A,B,C=1,2,......7)\label{eq:2}\end{eqnarray}
The structure constants $f_{ABC}$ are completely antisymmetric and
take the value $1$ i.e. $f_{ABC}=+1=(123),\,(471),\,(257),\,(165),\,(624),\,(543),\,(736)$.
Here the octonion algebra $\mathcal{O}$ is described over the algebra
of rational numbers having the vector space of dimension $8$. Octonion
algebra is non associative and multiplication rules for its basis
elements given by equations (\ref{eq:2},\ref{eq:3}) are then generalized
in the following table:

\begin{tabular}{|c||c||c||c||c||c||c||c|}
\hline 
$\cdot$ & $e_{1}$ & $e_{2}$ & $e_{3}$ & $e_{4}$ & $e_{5}$ & $e_{6}$ & $e_{7}$\tabularnewline
\hline
\hline 
$e_{1}$ & $-1$ & $e_{3}$ & $-e_{2}$ & $e_{7}$ & $-e_{6}$ & $e_{5}$ & $-e_{4}$\tabularnewline
\hline
\hline 
$e_{2}$ & $-e_{3}$ & $-1$ & $e_{1}$ & $e_{6}$ & $e_{7}$ & $-e_{4}$ & $-e_{5}$\tabularnewline
\hline
\hline 
$e_{3}$ & $e_{2}$ & $-e_{1}$ & $-1$ & $-e_{5}$ & $e_{4}$ & $e_{7}$ & $-e_{6}$\tabularnewline
\hline
\hline 
$e_{4}$ & $-e_{7}$ & $-e_{6}$ & $e_{5}$ & $-1$ & $-e_{3}$ & $e_{2}$ & $e_{1}$\tabularnewline
\hline
\hline 
$e_{5}$ & $e_{6}$ & $-e_{7}$ & $-e_{4}$ & $e_{3}$ & $-1$ & $-e_{1}$ & $e_{2}$\tabularnewline
\hline
\hline 
$e_{6}$ & $-e_{5}$ & $e_{4}$ & $-e_{7}$ & $-e_{2}$ & $e_{1}$ & $-1$ & $e_{3}$\tabularnewline
\hline
\hline 
$e_{7}$ & $e_{4}$ & $e_{5}$ & $e_{6}$ & $-e_{1}$ & $-e_{2}$ & $-e_{3}$ & $-1$\tabularnewline
\hline
\end{tabular}

\begin{description}
\item [{~~~~~~~~~~~~~~Table1-}] Octonion Multiplication table
\end{description}
Hence, we get the following relations among octonion basis elements
i.e.

\begin{eqnarray}
\left[e_{A},\,\, e_{B}\right] & = & 2f_{ABC}e_{C};\,\,\,\,\,\,\,\left\{ e_{A},\,\, e_{B}\right\} =-\delta_{AB}e_{0};\,\,\,\,\, e_{A}(e_{B}e_{C})\neq(e_{A}e_{B})e_{C};\label{eq:3}\end{eqnarray}
where brackets $[\,\,]$ and $\{\,\,\}$ are used respectively for
commutation and the anti commutation relations while $\delta_{AB}$
is the usual Kronecker delta-Dirac symbol.Octonion conjugate is thus
defined as,

\begin{eqnarray}
\bar{x} & = & x_{0}e_{0}-x_{1}e_{1}-x_{2}e_{2}-x_{3}e_{3}-x_{4}e_{4}-x_{5}e_{5}-x_{6}e_{6}-x_{7}e_{7}\nonumber \\
= & x_{0}e_{0} & -\sum_{A=1}^{7}x_{A}e_{A}\,\,\,\,\,\,\,\,\,\,\,(A=1,2,.....,7).\label{eq:4}\end{eqnarray}
An Octonion can be decomposed in terms of its scalar $(Sc(x))$ and
vector $(Vec(x))$ parts as 

\begin{eqnarray}
Sc(x) & = & \frac{1}{2}(x+\bar{x})=x_{0};\,\,\,\,\,\,\, Vec(x)=\frac{1}{2}(x-\bar{x})=\sum_{A=1}^{7}x_{A}e_{A}\label{eq:5}\end{eqnarray}
Conjugates of product of two octonions and its own are described as

\begin{eqnarray}
(\overline{xy}) & = & \overline{y}\,\overline{x}\,\,\,;\,\,\,\,\,\,\,\overline{(\bar{x})}\,\,=x\label{eq:6}\end{eqnarray}
while the scalar product of two octonions is defined as 

\begin{eqnarray}
\left\langle x\,,\, y\,\right\rangle  & =\sum_{\alpha=0}^{7} & x_{\alpha}y_{\alpha}=\frac{1}{2}(x\,\bar{y}+y\,\bar{x})=\frac{1}{2}(\bar{x}\, y+\bar{y}\, x)\label{eq:7}\end{eqnarray}
which can be written in terms of octonion units as

\begin{eqnarray}
\left\langle e_{A}\,,\, e_{B}\,\right\rangle  & = & \frac{1}{2}(e_{A}\overline{e_{B}}+e_{B}\overline{e_{A}})=\frac{1}{2}(\overline{e_{A}}e_{B}+\overline{e_{B}}e_{A})=\delta_{AB}.\label{eq:8}\end{eqnarray}
The norm of the octonion $N(x)$ is defined as

\begin{eqnarray}
N(x)=\overline{x}x & =x\,\bar{x} & =\sum_{\alpha=0}^{7}x_{\alpha}^{2}e_{0}\label{eq:9}\end{eqnarray}
which is zero if $x=0$, and is always positive otherwise. It also
satisfies the following property of normed algebra

\begin{eqnarray}
N(xy) & =N(x)N(y) & =N(y)N(x).\label{eq:10}\end{eqnarray}
As such, for a nonzero octonion $x$ , we define its inverse as

\begin{eqnarray}
x^{-1} & = & \frac{\bar{x}}{N(x)}\label{eq:11}\end{eqnarray}
which shows that

\begin{eqnarray}
x^{-1}x & =xx^{-1} & =1.e_{0};\,\,\,\,(xy)^{-1}=y^{-1}x^{-1}.\label{eq:12}\end{eqnarray}

\section{Octonion Wave Equation}

A lot of literature \cite{key-17,key-18,key-19,key-20} has already
been available on octonion wave equation. Accordingly, let us define
the octonion differential operator $D$ as

\begin{eqnarray}
D & = & \sum_{\mu=0}^{7}e_{\mu}D_{\mu},\label{eq:13}\end{eqnarray}
where $D_{\mu}$ are described as the components of a differential
operator in an eight dimensional representation. We describe a function
of an octonion variable as\begin{eqnarray}
\mathcal{F}(X) & =\sum_{\mu=0}^{7} & e_{\mu}f_{\mu}(X)=f_{0}+e_{1}f_{1}+e_{2}f_{2}+.....+e_{7}f_{7},\label{eq:14}\end{eqnarray}
where $f_{\mu}$ are scalar functions. Since octonions are neither
commutative nor associative, one has to be very careful to multiply
the octonion either from left or from right in terms of regularity
conditions \cite{key-17}. As such, a function $\mathcal{F}(X)$ of
an octonion variable $X={\displaystyle \sum_{\mu=0}^{7}e_{\mu}X_{\mu}}$
is left regular at $X$ if and only if $\mathcal{F}(X)$ satisfies
the condition

\begin{eqnarray}
D\mathcal{F}(X) & = & 0.\label{eq:15}\end{eqnarray}
Similarly, a function $G(X)$ is a right regular if and only if

\begin{eqnarray}
G(X)D & = & 0,\label{eq:16}\end{eqnarray}
where $G(X)=g_{0}+g_{1}e_{1}+g_{2}e_{2}+.....+g_{7}e_{7}.$ Then we
get 

\begin{eqnarray}
D\mathcal{F} & =I= & I_{0}+I_{1}e_{1}+I_{2}e_{2}+I_{3}e_{3}+I_{4}e_{4}+I_{5}e_{5}+I_{6}e_{6}+I_{7}e_{7},\label{eq:17}\end{eqnarray}
where

\begin{eqnarray}
I_{0} & = & \partial_{0}f_{0}-\partial_{1}f_{1}-\partial_{2}f_{2}-\partial_{3}f_{3}-\partial_{4}f_{4}-\partial_{5}f_{5}-\partial_{6}f_{6}-\partial_{7}f_{7};\nonumber \\
I_{1} & = & \partial_{0}f_{1}+\partial_{1}f_{0}+\partial_{2}f_{3}-\partial_{3}f_{2}+\partial_{6}f_{5}-\partial_{5}f_{6}-\partial_{7}f_{4}+\partial_{4}f_{7};\nonumber \\
I_{2} & = & \partial_{0}f_{2}+\partial_{2}f_{0}+\partial_{3}f_{1}-\partial_{1}f_{3}+\partial_{4}f_{6}-\partial_{6}f_{4}-\partial_{7}f_{5}+\partial_{5}f_{7};\nonumber \\
I_{3} & = & \partial_{0}f_{3}+\partial_{3}f_{0}+\partial_{1}f_{2}-\partial_{2}f_{1}+\partial_{6}f_{7}-\partial_{7}f_{6}+\partial_{5}f_{4}-\partial_{4}f_{5};\nonumber \\
I_{4} & = & \partial_{0}f_{4}+\partial_{4}f_{0}+\partial_{3}f_{5}-\partial_{5}f_{3}-\partial_{2}f_{6}+\partial_{6}f_{2}-\partial_{1}f_{7}+\partial_{7}f_{1};\nonumber \\
I_{5} & = & \partial_{0}f_{5}+\partial_{5}f_{0}+\partial_{1}f_{6}-\partial_{6}f_{1}+\partial_{7}f_{2}-\partial_{2}f_{7}-\partial_{3}f_{4}+\partial_{4}f_{3};\nonumber \\
I_{6} & = & \partial_{0}f_{6}+\partial_{6}f_{0}-\partial_{1}f_{5}+\partial_{5}f_{1}+\partial_{2}f_{4}-\partial_{4}f_{2}-\partial_{3}f_{7}+\partial_{7}f_{3};\nonumber \\
I_{7} & = & \partial_{0}f_{7}+\partial_{7}f_{0}+\partial_{1}f_{4}-\partial_{4}f_{1}+\partial_{2}f_{5}-\partial_{5}f_{2}-\partial_{6}f_{3}+\partial_{3}f_{6}.\label{eq:18}\end{eqnarray}
The regularity condition (\ref{eq:15}) may now be considered as a
homogeneous octonion wave equation for octonion variables without
sources. On the other hand, equation (\ref{eq:17}) is considered
as the inhomogeneous wave equation $D\mathcal{F}=I$.

\section{Octonion Formulation for Generalized Fields of Dyons}

In order to write the various quantum equations of dyons in octonion
formulation, let us start with potential octonion

\begin{eqnarray}
\mathbb{V} & = & e_{0}V_{0}+e_{1}V_{1}+e_{2}V_{2}+e_{3}V_{3}+e_{4}V_{4}+e_{5}V_{5}+e_{6}V_{6}+e_{7}V_{7}\label{eq:19}\end{eqnarray}
and we identify its components as 

\begin{equation}
(V_{0},V_{1},V_{2},V_{3},V_{4},V_{5},V_{6},V_{7})\Longrightarrow(\varphi,\, A_{x},\, A_{y},A_{z},\, iB_{x},\, iB_{y},\, iB_{z},i\phi)\,\,\,\,(i=\sqrt{-1})\label{eq:20}\end{equation}
where $(\phi,\, A_{x},\, A_{y},\, A_{z})=(\phi,\overrightarrow{A}\,)=\left\{ A_{\mu}\right\} $
and $(\varphi,\, B_{x},\, B_{y},\, B_{z})=(\varphi,\overrightarrow{B}\,)=\left\{ B_{\mu}\right\} $
are respectively described as the components of electric $\left\{ A_{\mu}\right\} $
and magnetic $\left\{ B_{\mu}\right\} $ four potentials of dyons
(particles carrying simultaneously the electric and magnetic charges).
Equation (\ref{eq:19}) may then be written as 

\begin{equation}
\mathbb{V=}e_{1}(A_{x}+ie_{7}B_{x})+e_{2}(A_{y}+ie_{7}B_{y})+e_{3}(A_{z}+ie_{7}B_{z})+(\varphi+ie_{7}\phi)=e_{1}\mathrm{V}_{\mathrm{x}}+e_{2}\mathrm{V}_{\mathrm{y}}+e_{3}\mathrm{V_{z}}+ie_{7}\emptyset\label{eq:21}\end{equation}
where $(\emptyset,\mathrm{V_{x},}\mathrm{V_{y},\mathrm{V_{z}}})=(\emptyset,\overrightarrow{\mathrm{V}}\,)=\left\{ \mathrm{V_{\mu}}\right\} $
are then be described as the components of generalized four potential
$\left\{ \mathrm{V_{\mu}}\right\} $ associated with generalized charge
$(q=e+i\, g$) (where $e$ and $g$ are respectively known as electric
and magnetic charges) of dyons \cite{key-6}. In order to obtain the
generalized field equations of dyons in four dimensional space time,
we identify differential operator (\ref{eq:13}) to be four dimensional
and hence we may write equation (\ref{eq:13}) as 

\begin{eqnarray}
D\longmapsto\boxdot & = & e_{1}\frac{\partial}{\partial x}+e_{2}\frac{\partial}{\partial y}+e_{3}\frac{\partial}{\partial z}-ie_{7}\frac{\partial}{\partial t}\label{eq:22}\end{eqnarray}
where we have taken other components like $\partial_{0},\partial_{4},\partial_{5},\partial_{6}$of
equation (\ref{eq:13}) vanishing. Octonion conjugate of equation
(\ref{eq:22}) may then be written as 

\begin{eqnarray}
\overline{\boxdot} & = & -e_{1}\frac{\partial}{\partial x}-e_{2}\frac{\partial}{\partial y}-e_{3}\frac{\partial}{\partial z}+ie_{7}\frac{\partial}{\partial t}.\label{eq:23}\end{eqnarray}
Now operating $\overline{\boxdot}$ given by equation (\ref{eq:23})
to octonion potential $\mathtt{\mathrm{\mathcal{\mathsf{\mathbf{\mathcal{\mathfrak{\mathbb{V}}}}}}}}$of
equation (\ref{eq:21}), we get 

\begin{align}
\overline{\boxdot}\,\mathbb{V=} & -e_{0}(\overrightarrow{\nabla}\cdot\overrightarrow{A}+\frac{\partial\phi}{\partial t})\nonumber \\
+ & e_{1}(-\frac{\partial\varphi}{\partial x}+\frac{\partial A_{z}}{\partial y}-\frac{\partial A_{y}}{\partial z}-\frac{\partial B_{x}}{\partial t})\nonumber \\
+ & e_{2}(-\frac{\partial\varphi}{\partial y}+\frac{\partial A_{x}}{\partial z}-\frac{\partial A_{z}}{\partial zx}-\frac{\partial B_{y}}{\partial t})\nonumber \\
+ & e_{3}(-\frac{\partial\varphi}{\partial z}+\frac{\partial A_{y}}{\partial x}-\frac{\partial A_{x}}{\partial y}-\frac{\partial B_{z}}{\partial t})\nonumber \\
- & ie_{4}(-\frac{\partial\phi}{\partial x}-\frac{\partial B_{z}}{\partial y}+\frac{\partial B_{y}}{\partial z}-\frac{\partial A_{x}}{\partial t})\nonumber \\
- & ie_{5}(-\frac{\partial\phi}{\partial y}-\frac{\partial B_{x}}{\partial z}+\frac{\partial B_{z}}{\partial x}-\frac{\partial A_{y}}{\partial t})\nonumber \\
- & ie_{6}(-\frac{\partial\phi}{\partial z}-\frac{\partial B_{y}}{\partial x}+\frac{\partial B_{x}}{\partial y}-\frac{\partial A_{z}}{\partial t})\nonumber \\
+ & ie_{7}(\overrightarrow{\nabla}\cdot\overrightarrow{B}+\frac{\partial\varphi}{\partial t}).\label{eq:24}\end{align}
We are using S.I. system of natural units $(c=\hbar=1)$. On applying
the Lorentz gauge conditions, respectively for the dynamics of electric
and magnetic charges of dyons as 

\begin{eqnarray}
\overrightarrow{\nabla}\cdot\overrightarrow{A}+\frac{\partial\phi}{\partial t} & = & 0;\nonumber \\
\overrightarrow{\nabla}\cdot\overrightarrow{B}+\frac{\partial\varphi}{\partial t} & = & 0,\label{eq:25}\end{eqnarray}
we find the following octonion form of equation (\ref{eq:24}) i.e. 

\begin{eqnarray}
\overline{\boxdot}\,\mathbb{V} & = & \mathbb{F}\label{eq:26}\end{eqnarray}
where $\mathbb{F}$ is again an octonion reproduces the generalized
electromagnetic fields of dyons. It is thus described by 

\begin{eqnarray}
\mathbb{F} & = & e_{0}F_{0}+e_{1}F_{1}+e_{2}F_{2}+e_{3}F_{3}+e_{4}F_{4}+e_{5}F_{5}+e_{6}F_{6}+e_{7}F_{7}\label{eq:27}\end{eqnarray}
where 

\begin{eqnarray}
F_{0} & = & -(\overrightarrow{\nabla}\cdot\overrightarrow{A}+\frac{\partial\phi}{\partial t})=0;\nonumber \\
F_{1} & = & (-\frac{\partial\varphi}{\partial x}+\frac{\partial A_{z}}{\partial y}-\frac{\partial A_{y}}{\partial z}-\frac{\partial B_{x}}{\partial t})=H_{x}\nonumber \\
F_{2} & = & (-\frac{\partial\varphi}{\partial y}+\frac{\partial A_{x}}{\partial z}-\frac{\partial A_{z}}{\partial x}-\frac{\partial B_{y}}{\partial t})=H_{y}\nonumber \\
F_{3} & = & (-\frac{\partial\varphi}{\partial z}+\frac{\partial A_{y}}{\partial x}-\frac{\partial A_{x}}{\partial y}-\frac{\partial B_{z}}{\partial t})=H_{z}\nonumber \\
F_{4} & = & -i\,(-\frac{\partial\phi}{\partial x}-\frac{\partial B_{z}}{\partial y}+\frac{\partial B_{y}}{\partial z}-\frac{\partial A_{x}}{\partial t})=-i\, E_{x}\nonumber \\
F_{5} & = & -i\,(-\frac{\partial\phi}{\partial y}-\frac{\partial B_{x}}{\partial z}+\frac{\partial B_{z}}{\partial x}-\frac{\partial A_{y}}{\partial t})=-i\, E_{y}\nonumber \\
F_{6} & = & -i\,(-\frac{\partial\phi}{\partial z}-\frac{\partial B_{y}}{\partial x}+\frac{\partial B_{x}}{\partial y}-\frac{\partial A_{z}}{\partial t})=-i\, E_{z}\nonumber \\
F_{7} & = & i\,(\overrightarrow{\nabla}\cdot\overrightarrow{B}+\frac{\partial\varphi}{\partial t})=0.\label{eq:28}\end{eqnarray}
Let us define \cite{key-6,key-21} the generalized electric $(\overrightarrow{E})$
and magnetic $(\overrightarrow{H})$ fields of dyons in terms of components
of electric and magnetic four potentials as

\begin{eqnarray}
\overrightarrow{E} & =- & \frac{\partial\overrightarrow{A}}{\partial t}-\overrightarrow{\nabla}\phi-\overrightarrow{\nabla}\times\overrightarrow{B};\nonumber \\
\overrightarrow{H} & = & -\frac{\partial\overrightarrow{B}}{\partial t}-\overrightarrow{\nabla}\varphi+\overrightarrow{\nabla}\times\overrightarrow{A}.\label{eq:29}\end{eqnarray}
Thus equation (\ref{eq:27}) reduces to 

\begin{align}
\mathbb{F=} & e_{1}(H_{x}+ie_{7}E_{x})+e_{2}(H_{y}+ie_{7}E_{y})+e_{3}(H_{z}+ie_{7}E_{z})\nonumber \\
= & e_{1}\Psi_{x}+e_{2}\Psi_{y}+e_{3}\Psi_{z}\label{eq:30}\end{align}
where $\overrightarrow{\Psi}=\overrightarrow{H}+i\, e_{7}\overrightarrow{E}$
is the generalized vector field \cite{key-6,key-21} of dyons. Now
applying the differential operator (\ref{eq:22}) to equation (\ref{eq:30}),
we get

\begin{align}
\boxdot\,\mathbb{\mathbb{F}=} & -e_{0}(\frac{\partial H_{x}}{\partial x}+\frac{\partial H_{y}}{\partial y}+\frac{\partial H_{z}}{\partial z})\nonumber \\
+ & e_{1}(\frac{\partial H_{z}}{\partial y}-\frac{\partial H_{y}}{\partial z}-\frac{\partial E_{x}}{\partial t})\nonumber \\
+ & e_{2}(\frac{\partial H_{x}}{\partial z}-\frac{\partial H_{z}}{\partial x}-\frac{\partial E_{y}}{\partial t})\nonumber \\
+ & e_{3}(\frac{\partial H_{y}}{\partial x}-\frac{\partial H_{x}}{\partial y}-\frac{\partial E_{z}}{\partial t})\nonumber \\
+ & ie_{4}(\frac{\partial E_{y}}{\partial z}-\frac{\partial E_{z}}{\partial y}-\frac{\partial H_{x}}{\partial t})\nonumber \\
+ & ie_{5}(\frac{\partial E_{z}}{\partial x}-\frac{\partial E_{x}}{\partial z}-\frac{\partial H_{y}}{\partial t})\nonumber \\
+ & ie_{6}(\frac{\partial E_{x}}{\partial y}-\frac{\partial E_{y}}{\partial x}-\frac{\partial H_{z}}{\partial t})\nonumber \\
+ & ie_{7}(\frac{\partial E_{x}}{\partial x}+\frac{\partial E_{y}}{\partial y}+\frac{\partial E_{z}}{\partial z}).\label{eq:31}\end{align}
or 

\begin{align}
\mathbb{\boxdot\mathbb{F=}} & -e_{0}(\overrightarrow{\nabla}.\overrightarrow{H})\nonumber \\
+ & e_{1}[(\overrightarrow{\nabla}\times\overrightarrow{H})_{x}-\frac{\partial E_{x}}{\partial t}]\nonumber \\
+ & e_{2}[(\overrightarrow{\nabla}\times\overrightarrow{H})_{y}-\frac{\partial E_{y}}{\partial t}]\nonumber \\
+ & e_{3}[(\overrightarrow{\nabla}\times\overrightarrow{H})_{z}-\frac{\partial E_{z}}{\partial t}]\nonumber \\
+ & ie_{4}[(\overrightarrow{\nabla}\times\overrightarrow{E})_{x}-\frac{\partial H_{x}}{\partial t}]\nonumber \\
+ & ie_{5}[(\overrightarrow{\nabla}\times\overrightarrow{E})_{y}-\frac{\partial H_{y}}{\partial t}]\nonumber \\
+ & ie_{6}[(\overrightarrow{\nabla}\times\overrightarrow{E})_{z}-\frac{\partial H_{z}}{\partial t}]\nonumber \\
+ & ie_{7}(\overrightarrow{\nabla}.\overrightarrow{E})\label{eq:32}\end{align}
These equations (\ref{eq:31},\ref{eq:32}) may then be written in
following compact notation in terms of an octonion i.e.

\begin{eqnarray}
\boxdot\mathbb{F} & = & \mathbb{J}\label{eq:33}\end{eqnarray}
where $\mathbb{J}$ is again an octonion and is re-described as the
Octonion form of generalized current given by

\begin{align}
\mathbb{J=}-e_{0}\varrho & \mathrm{+e_{1}\mathrm{\mathrm{j}}_{\mathrm{x}}+e_{2}\mathrm{j}_{\mathrm{y}}+e_{3}\mathrm{j}_{\mathrm{z}}-ie_{4}}\mathrm{k_{x}-ie_{5}\mathrm{k_{y}-ie_{6}\mathrm{k_{z}}+ie_{7}\mathrm{\rho}}}\nonumber \\
= & \mathrm{(e_{1}\mathrm{\mathrm{j}}_{\mathrm{x}}+e_{2}\mathrm{j}_{\mathrm{y}}+e_{3}\mathrm{j}_{\mathrm{z}}--e_{0}\varrho})+i(\mathrm{e}_{1}\mathrm{k_{x}+e_{2}\mathrm{k_{y}+e_{3}\mathrm{k_{z}}-\mathrm{\rho}}})\mathrm{e_{7}}\nonumber \\
= & \mathrm{e_{1}}(\mathrm{\mathrm{j}}_{\mathrm{x}}+i\mathrm{e_{7}k_{x})+\mathrm{e_{2}(j_{y}+ie_{7}k_{z})+e_{3}(j_{z}+k_{z})-(\rho+ie_{7}\varrho)}}\nonumber \\
= & \mathrm{e_{1}}\mathrm{J_{x}}+\mathrm{e_{2}J_{y}+e_{3}J_{z}+e_{0}J_{0}}\label{eq:34}\end{align}
where $\mathrm{(\rho,\,\overrightarrow{j})=\{\mathrm{j_{\mu}}\}}$,
$(\varrho,\,\overrightarrow{j})=\{\mathrm{k_{\mu}}\}$ and $(\mathrm{J_{0},\overrightarrow{J})=\{\mathrm{J_{\mu}}\}}$
are respectively the four currents associated with electric charge,
magnetic monopole and generalized fields of dyons. Equations (\ref{eq:31}-\ref{eq:34})
thus lead to following differential equations 

\begin{eqnarray}
(\overrightarrow{\nabla}\cdot\overrightarrow{H}) & = & \varrho\nonumber \\
(\overrightarrow{\nabla}\times\overrightarrow{H})_{x} & = & \frac{\partial E_{x}}{\partial t}+\mathrm{j_{x}}\nonumber \\
(\overrightarrow{\nabla}\times\overrightarrow{H})_{y} & = & \frac{\partial E_{y}}{\partial t}+\mathrm{j_{y}}\nonumber \\
(\overrightarrow{\nabla}\times\overrightarrow{H})_{z} & = & \frac{\partial E_{z}}{\partial t}+\mathrm{j_{z}}\nonumber \\
(\overrightarrow{\nabla}\times\overrightarrow{E})_{x} & =- & \frac{\partial H_{x}}{\partial t}-\mathrm{k_{x}}\nonumber \\
(\overrightarrow{\nabla}\times\overrightarrow{E})_{y} & =- & \frac{\partial H_{y}}{\partial t}-\mathrm{k_{y}}\nonumber \\
(\overrightarrow{\nabla}\times\overrightarrow{E})_{z} & =- & \frac{\partial H_{z}}{\partial t}-\mathrm{k_{z}}\nonumber \\
(\overrightarrow{\nabla}\cdot\overrightarrow{E}) & = & \rho.\label{eq:35}\end{eqnarray}
Equation (\ref{eq:35}) may then be written as 

\begin{eqnarray}
(\overrightarrow{\nabla}\cdot\overrightarrow{E}) & = & \rho\nonumber \\
(\overrightarrow{\nabla}\times\overrightarrow{E}) & =- & \frac{\partial\overrightarrow{H}}{\partial t}-\mathrm{\overrightarrow{k}}\nonumber \\
(\overrightarrow{\nabla}\times\overrightarrow{H}) & = & \frac{\partial\overrightarrow{E}}{\partial t}+\mathrm{\overrightarrow{j}}\nonumber \\
(\overrightarrow{\nabla}\cdot\overrightarrow{H}) & = & \varrho\label{eq:36}\end{eqnarray}
which are the generalized Dirac-Maxwell's equations of generalized
fields of dyons \cite{key-21,key-22}.

\begin{align}
\boxdot\bar{\boxdot} & \mathbb{V}=\mathbb{F}\nonumber \\
= & e_{0}(\frac{\partial^{2}\varphi}{\partial x^{2}}+\frac{\partial^{2}\varphi}{\partial y^{2}}+\frac{\partial^{2}\varphi}{\partial z^{2}}+\frac{\partial^{2}B_{x}}{\partial x\partial t}+\frac{\partial^{2}B_{y}}{\partial y\partial t}+\frac{\partial^{2}B_{z}}{\partial z\partial t})\nonumber \\
- & e_{1}(\frac{\partial^{2}A_{x}}{\partial y^{2}}+\frac{\partial^{2}A_{x}}{\partial z^{2}}-\frac{\partial^{2}A_{x}}{\partial t^{2}}-\frac{\partial^{2}A_{z}}{\partial x\partial z}-\frac{\partial^{2}A_{y}}{\partial y\partial x}-\frac{\partial^{2}\phi}{\partial x\partial t})\nonumber \\
- & e_{2}(\frac{\partial^{2}A_{y}}{\partial x^{2}}+\frac{\partial^{2}A_{y}}{\partial z^{2}}-\frac{\partial^{2}A_{y}}{\partial t^{2}}-\frac{\partial^{2}A_{x}}{\partial y\partial x}-\frac{\partial^{2}A_{z}}{\partial y\partial z}-\frac{\partial^{2}\phi}{\partial y\partial t})\nonumber \\
- & e_{3}(\frac{\partial^{2}A_{z}}{\partial z^{2}}+\frac{\partial^{2}A_{z}}{\partial y^{2}}-\frac{\partial^{2}A_{z}}{\partial t^{2}}-\frac{\partial^{2}A_{y}}{\partial z\partial y}-\frac{\partial^{2}A_{x}}{\partial z\partial x}-\frac{\partial^{2}\phi}{\partial z\partial t})\nonumber \\
+ & ie_{4}(-\frac{\partial^{2}B_{x}}{\partial y^{2}}-\frac{\partial^{2}B_{x}}{\partial z^{2}}+\frac{\partial^{2}B_{x}}{\partial t^{2}}+\frac{\partial^{2}B_{y}}{\partial x\partial y}+\frac{\partial^{2}B_{z}}{\partial z\partial x}+\frac{\partial^{2}\varphi}{\partial x\partial t})\nonumber \\
+ & ie_{5}(-\frac{\partial^{2}B_{y}}{\partial x^{2}}-\frac{\partial^{2}B_{y}}{\partial z^{2}}+\frac{\partial^{2}B_{y}}{\partial t^{2}}+\frac{\partial^{2}B_{z}}{\partial z\partial y}+\frac{\partial^{2}B_{x}}{\partial y\partial x}+\frac{\partial^{2}\varphi}{\partial y\partial t})\nonumber \\
+ & ie_{6}(-\frac{\partial^{2}B_{z}}{\partial x^{2}}-\frac{\partial^{2}B_{z}}{\partial y^{2}}+\frac{\partial^{2}B_{z}}{\partial t^{2}}+\frac{\partial^{2}B_{x}}{\partial x\partial z}+\frac{\partial^{2}B_{y}}{\partial z\partial y}+\frac{\partial^{2}\varphi}{\partial z\partial t})\nonumber \\
- & ie_{7}(\frac{\partial^{2}\phi}{\partial x^{2}}+\frac{\partial^{2}\phi}{\partial y^{2}}+\frac{\partial^{2}\phi}{\partial z^{2}}+\frac{\partial^{2}A_{x}}{\partial x\partial t}+\frac{\partial^{2}A_{y}}{\partial t\partial y}+\frac{\partial^{2}A_{z}}{\partial z\partial t}).\label{eq:37}\end{align}
Equation (\ref{eq:37}) then reduces to\begin{eqnarray}
\bar{\boxdot\boxdot\mathbb{V}} & =\bar{\boxdot}\boxdot\mathbb{V\mathrm{\mathbb{=}}} & \mathbb{J}\label{eq:38}\end{eqnarray}
where $\mathbb{J}$ is described as the octonion form of generalized
current associated with dyon and is already given by equation (\ref{eq:33}).
Equation (\ref{eq:37}) may also be written as 

\begin{align}
\bar{\boxdot\boxdot\mathbb{V}}=\bar{\boxdot}\boxdot\mathbb{V=} & e_{0}[\nabla^{2}\varphi+\frac{\partial}{\partial t}(\overrightarrow{\nabla}\cdot\overrightarrow{B})]\nonumber \\
- & e_{1}[\mathrm{\mathtt{\square}}A_{x}-\frac{\partial}{\partial x}(\overrightarrow{\nabla}\cdot\overrightarrow{A}+\frac{\partial\phi}{\partial t})]\nonumber \\
- & e_{2}[\mathrm{\mathtt{\square}}A_{y}-\frac{\partial}{\partial y}(\overrightarrow{\nabla}\cdot\overrightarrow{A}+\frac{\partial\phi}{\partial t})]\nonumber \\
- & e_{3}[\mathrm{\mathtt{\square}}A_{z}-\frac{\partial}{\partial z}(\overrightarrow{\nabla}\cdot\overrightarrow{A}+\frac{\partial\phi}{\partial t})]\nonumber \\
-i & e_{4}[\mathrm{\mathtt{\square}}B_{x}-\frac{\partial}{\partial x}(\overrightarrow{\nabla}\cdot\overrightarrow{B}+\frac{\partial\varphi}{\partial t})]\nonumber \\
-i & e_{5}[\mathrm{\mathtt{\square}}B_{y}-\frac{\partial}{\partial y}(\overrightarrow{\nabla}\cdot\overrightarrow{B}+\frac{\partial\varphi}{\partial t})]\nonumber \\
-i & e_{6}[\mathrm{\mathtt{\square}}B_{z}-\frac{\partial}{\partial z}(\overrightarrow{\nabla}\cdot\overrightarrow{B}+\frac{\partial\varphi}{\partial t})]\nonumber \\
-i & e_{7}[\nabla^{2}\phi+\frac{\partial}{\partial t}(\overrightarrow{\nabla}\cdot\overrightarrow{A})],\label{eq:39}\end{align}
where $\nabla^{2}=\frac{\partial^{2}}{\partial x^{2}}+\frac{\partial^{2}}{\partial y^{2}}+\frac{\partial^{2}}{\partial z^{2}}$
and $\mathtt{\square}=\frac{\partial^{2}}{\partial x^{2}}+\frac{\partial^{2}}{\partial y^{2}}+\frac{\partial^{2}}{\partial z^{2}}$-$\frac{\partial^{2}}{\partial t^{2}}=\nabla^{2}-\frac{\partial^{2}}{\partial t^{2}}$
. Using the Lorentz gauge conditions (\ref{eq:25}) and the definition
of octonion valued generalized current of dyon given by equation (\ref{eq:34}),
we get 

\begin{eqnarray}
\mathrm{\mathtt{\square}}\phi & = & \rho;\,\,\,\,\,\,\,\,\,\,\,\mathrm{\mathtt{\square}}\varphi=\varrho;\nonumber \\
\mathrm{\mathtt{\square}}\mathrm{A_{\mu}} & = & \mathrm{j_{\mu};\,\,\,\,\,\,\,\,\,\,\mathrm{\mathtt{\square}}B_{\mu}=k_{\mu}}.\label{eq:40}\end{eqnarray}
As such, we have obtained consistently the generalized Dirac Maxwell's
(GDM) equations from octonion wave equations on considering the non
associativity of octonion variables. The advantages of present formalism
are discussed in terms of compact and simpler notations of octonion
valued potential, field and currents of dyons despite of non associativity
of octonions. The present octonion reformulation of generalized fields
of dyons represents well the invariance of field equations under Lorentz
and duality transformations. It also reproduces the dynamics of electric
(magnetic) charge yielding to the usual form of Maxwell's equations
in the absence of magnetic (electric charge) in compact, simpler and
consistent way. Octonion element $e_{7}$ has been considered to be
invariant and the theory resembles with the bi-quaternion theory of
generalized fields of dyons \cite{key-9}. In the forthcoming paper
theory of split octonion will be discussed.

\end{document}